\documentclass{aa}
\usepackage{graphicx}

\def\msun{M$_{\odot}$}

\def\lsun{L$_{\odot}$}

\def\sbs{SBS\,1150+599A}
\begin{document}

     \thesaurus{06     
                (09.16.1;  
                 09.16.2;  
                 09.08.1;  
                 11.08.1;  
                 08.02.5;  
                 08.14.2)} 

     \title{SBS~1150+599A: an extremely oxygen-poor planetary nebula in
the Galactic halo?}

     \author{Gaghik\,H.~Tovmassian
         \inst{1}
         \and Gra\.{z}yna~Stasi\'{n}ska
         \inst{2}
         \and Vahram\,H.~Chavushyan
         \inst{3}
         \and Sergei\,V.~Zharikov
         \inst{1}\thanks
         {Special Astrophys. Obs., 357147 Nizhnij Arkhyz,
         Russia}
         \and Carlos~Gutierrez 
         \inst{4}
         \and Francisco~Prada
         \inst{1}\thanks{currently Centro Astronomico Hispano-Aleman
                          Apdo 511, E-04080 Almeria, Spain}
            }

     \offprints{G.\,Tovmassian}

     \institute{OAN, Instituto de Astronom\'{\i}a, UNAM, M\'{e}xico\\
                email: gag@astrosen.unam.mx
         \and
              DAEC, Observatoire de Meudon, 92195 Meudon Cedex, France\\
                email: grazyna.stasinska@obspm.fr
         \and
              Instituto Nacional de Astrof\'{\i}sica
   Optica y Electr\'onica, AP 51 y 216, Puebla, Pue., M\'{e}xico\\
              email: vahram@inaoep.mx
           \and
               Instituto de Astrof\'{\i}sica de Canarias,
Ave. Via L\'actea s/n, 38200 La Laguna, Tenerife, Spain\\
               email: cgc@ll.iac.es
               }

     \date{Received ; accepted }

\authorrunning {Tovmassian, Stasi\'{n}ska et\,al.}
\titlerunning {SBS~1150+599A}
\maketitle

     \markboth {G.H.~Tovmassian et al.}{ SBS1150+599A}

     \begin{abstract}

We report results of a spectrophotometric study of SBS\,1150+599A and
discuss the nature of this object based upon our data.  Our study
shows that SBS\,1150+599A is most probably
a planetary nebula located in the Galactic halo and
not a cataclysmic variable as originally
proposed by the authors of the Second Byurakan Survey from low
resolution spectroscopy.
We have further elaborated on the properties of  SBS\,1150+599A 
(now becoming PN G135.9+55.9) with
tools used for planetary nebula analysis. Our photoionization models
show that, in order to match the observational constraints, the
oxygen abundance in the nebula is probably extremely low, around
1/500 solar, which is one order of magnitude lower than the most
oxygen-poor planetary nebulae known so far. This finding has strong
implications on our understanding of the formation of planetary
nebulae and of the evolution of the Galactic halo.

        \keywords{ISM: planetary nebulae: general -- ISM: planetary nebulae: 
                  individual: \sbs\ (PN G135.9+55.9), --
                  Galaxies: halos -- Stars: binaries: symbiotic --
                  Stars: cataclysmic variables
}
     \end{abstract}

%

\section{Introduction}

\sbs\, pertains to the list of stellar-like objects from the Second Byurakan
Survey (SBS) (\cite{smb97}; \cite{jvn99}).
The Second Byurakan Survey is a low resolution objective prism
survey with a limiting magnitude of B$\sim$19\fm5 (Stepanian, 1994;
Bicay et al 2000).
The observing technique and the selection criteria
for the SBS objects have been described by Markarian \& Stepanian 
(\cite{mrk83}), Markarian et al. (\cite{mrk87}), and Stepanian 
(\cite{jiv94a, jiv94b}).  
Worth mentioning is that the selection criteria of the SBS
include the presence of a strong UV continuum, emission lines, and/or peculiar
energy distribution as inferred from the objective prism spectra. These
criteria have been successful in selecting objects such as UVX galaxies
(Markarian galaxies), as well as a broad class of AGNs (Carrasco et
al. \cite{car98}; \cite{jvn99}). As a by product, a large number of peculiar
stars, WDs, composite and emission objects, hot subdwarfs and other types of
objects have been detected (\cite{smb97}; \cite{jvn99}).
Among them several new Cataclysmic Variables(CV) were discovered.

Here, we report on a  spectrophotometric study of \sbs\ in order to 
clarify its nature. This object is located at very high
 Galactic latitude: $l_{II}=135.96$, $b_{II}=55.94$. It was
 first thought to be a CV based upon the appearance of 
its spectrum.
It shows strong hydrogen and helium lines superposed upon a hot continuum,
a characteristic pattern for CVs. Later, in a short communication,
Garnavich (\cite{gar99}) claimed that \sbs\, is a symbiotic star
or a young planetary nebula.
Our study shows  that this object is
not a CV or symbiotic, but probably a peculiar and rather old 
halo Planetary Nebula (PN).

In Section 2, we present the data acquisition and reduction
processes, in Section 3 we derive the observational parameters for \sbs.
In Section 4 we discuss the various hypothesis about the nature of
\sbs, and in Section 5 we perform a photoionization modeling of the
object, bringing constraints on the oxygen abundance and temperature
of the exciting star. In Section 6 we discuss some implications of
our work.


\begin{table*}[th]
\caption{The Log of Spectral Observations}
\begin{tabular}{llccrrl}
        \noalign{\smallskip}
        \hline
        \noalign{\smallskip}
Date UT & JD & Telescope + Equip. & Wavelength & Number & Exp. & Site \\
       &    &                    &     Range           &  of spectra
& sec. &     \\
   \noalign{\smallskip}
   \hline
   \noalign{\smallskip}
1998 March 20 & 2450892 & 2.1m, B\&Ch+CCD2K & 3600--6200 & 10 & 900/1200 &SPM\\
1998 June 06& 2450971 & 2.5m, IDS+CCD TEK3 & 4600-5500  & 2 & 1200 & La Palma\\
2000 March 19 & 2451622 & 2.1, LFOSC+CCD & 4000-7170 & 2 & 1800 & Cananea\\
2000 April 11 & 2451645 & 2.1m, B\&Ch+CCD TEK1 & 3700--7500 & 3 & 1200 & SPM\\
2000 April 12 & 2451646 & 6.0m, UAGS+PMCCD  & 3600-8260  & 1 & 900 & SAO\\

         \noalign{\smallskip}
        \hline
      \end{tabular}
     \label{obstab}
     \end{table*}

\section {Data acquisition and reduction}

\sbs\, was observed with the 6\,m telescope of the Special
Astrophysical Observatory
of the Russian Academy of Sciences (SAO RAN) in 1986. 
The observations were carried
out with the SP-124 spectrograph equipped with a
1024-channel photon counting system (IPCS)  (Drabek et\,al. \cite{dra85}),
installed at the Nasmith\,{\small I} focus.
No flux calibration is available. Based upon that low resolution
spectrum, \sbs\, was classified
as a CV ({\cite{smb97}; \cite{jvn99}).

Recently we acquired a new spectrum of the object on the  6\,m telescope
with the
   UAGS spectrograph and a CCD detector. Although we obtained the necessary
spectrophotometric standards,
the sky conditions were poor for correct absolute flux measurements.
The spectrum was obtained in the 3600-8260\,\AA\, range with a spectral
resolution of 4.65\,\AA/pix which,  combined with the deployed slit width,
gives overall resolution of 10\,\AA\,FWHM.

We also obtained a set of spectra of  \sbs\, with the 2.1m telescope
of the Observatorio Astronomico Nacional de San Pedro Martir (OAN SPM)
on 20 of March 1998. The Boller \& Chivens spectrograph was set in the 4150 to
6750\,\AA\, range with a resolution of 1.28\,\AA/pix corresponding to
4.8\,\AA\,FWHM. The slit was
set to 1.8 arcsec. The seeing was poor.
It was not possible to obtain standard stars for calibration during
the same night.
We used an averaged sensitivity curve for the corresponding grating
and settings of
the spectrograph to calibrate the data from this night.

Additional spectra of \sbs\, were obtained at the  2.1m telescope of OAN SPM on
11 of April 2000.
This time we used a 300 l/mm grating with the same
spectrograph
in order to observe the object in the $\lambda$\,3700-7500\,\AA\, wavelength
range with 8.0\,\AA\,FWHM
resolution. During these observations, in order to cover a larger
fraction of the
object,  we  opened the slit up to 3.0 arcsec although
the seeing was good and did not exceed 1.5 arcsec. Two spectrophotometric
standards were observed along with the object
and an He-Ar lamp was used as in the previous observations for wavelength
calibration.

The object was also observed on one occasion with the 2.5 m Isaac Newton
Telescope (INT) of Observatorio de La Palma. The IDS  spectrograph
was centered around the  H$\beta$ region (4600--5500\,\AA). A spectral
resolution of 1.6\,\AA\, was reached as measured  from the  FWHM of arc lines.
A Cu-Ar+CuNe lamp was
used for wavelength calibration. No flux calibration is available.

We also acquired a low resolution spectrum  of \sbs\, at the 2.1\,m telescope
 at Cananea, M\'exico in March of 2000. The LFOSC spectrograph 
(Zickgraf et al. \cite{zi97})
was deployed to cover the $\lambda$\,4000--7100\,\AA\, wavelength range with
$\approx$\,13\,\AA\, FWHM resolution. The long slit was deployed
at the entrance of this multi-object spectrograph  3 arcsec wide as 
projected on the sky.

For none of the observations was the slit oriented along the parallactic
angle. Usually
it was simply in East-West direction. For the SPM observations, due to the faintness
of the object and absence of appropriate slit-viewing detector, pointing and
subsequent guiding were not perfect and might cause inaccuracy of flux
determinations.

Most of the observed data  were reduced using IRAF
packages after applying standard procedures for bias and flat field
corrections. The observations acquired at SAO were reduced using the 
MIDAS system.
The MIDAS cosmic rays cleaning procedure was also applied to the rest of the  data.

The Journal of observations is presented in Table \ref{obstab}.

We performed an extensive search in publicly available surveys and databases
for any additional measurements  at the location of \sbs. IRAS has no detection
of any IR source within several arcminutes, even among rejected faint sources.
No emission was detected at short wavelengths. Neither ROSAT RASS nor UV
surveys show any  emission within corresponding error boxes. Among
radio surveys
the closest  registered source  lies 6 arcminutes away from the object.

\section {Spectral analysis and  observational parameters}

The spectrum of the object displays a blue  continuum with
strong narrow emission lines of \ion{H}{i}, \ion{He}{ii} and
marginally detected
[\ion{O}{iii}]. No trace of \ion{He}{i} lines
or lines from singly ionized ions of O, N or S are detected.
In Fig. \ref{fig1} the low/medium resolution spectra of the
\sbs\, are presented at different epochs.

The spectrum in the upper panel is the best flux--calibrated
spectrum obtained at SPM with the wide slit 
(it is in reasonably good agreement with
the latest SAO and Cananea
spectra). 
In the inset of the upper panel
the sum of 7 spectra (total exposure of 8400 sec) from previous SPM
observations is presented.
It reaches a S/N ratio of 20 in the continuum around H$\alpha$.
The faint lines of [\ion{O}{iii}\,]$\lambda$\,5007\,\AA\, and
\ion{He}{ii}\,$\lambda$\,5411\,\AA\
appear clearly in that spectrum.

The shape of the continuum and of the line profiles
do not change from epoch to epoch. There is a small discrepancy
of absolute fluxes between spectra, but this could easily be a result
of inadequate observational conditions rather than variability. These leaves
open the question of the variability of the \sbs.

In the lower panels of  Fig. \ref{fig1} the  individual profiles of
H$\beta$ and
\ion{He}{ii} lines from the higher spectral resolution
spectrum obtained at INT  are shown. This spectrum,
obtained with 1.5\,\AA\, resolution,
demonstrates that the emission lines of the object are relatively narrow.
The measured
values of FWHM of H$\beta$ and \ion{He}{ii} are 2.0\,\AA\, and
1.85\,\AA\, respectively.

Another important spectroscopic feature is that the emission lines
of \sbs\, are displaced towards short wavelengths. This blue shift
shows some dispersion from line to line and also from epoch to epoch.

Before we consider the  measured fluxes,  it is necessary to mention that
the emission lines are spatially extended, which indicates
an extended object.  The edge to edge width of the line in the spatial direction
on the INT spectrum is $\approx 14$ pix, which corresponds to an
object size of $\approx$ 10\arcsec. A similar value is obtained from 
the SPM spectra.

Table \ref{tab2} lists the parameters of the emission
lines. Equivalent widths and relative fluxes were measured in all the acquired
spectra and we decided to present the whole range of obtained values,
which gives a good idea about the errors of measurement. The spread is
up to  35\% in some line fluxes, however relative numbers are usually
more consistent among the various spectra.
We found systematically lower values of the H$\alpha$/H$\beta$\,  ratio
in spectra obtained at SPM. We suspect  that there were problems with
the linearity of the CCD used in those observations and that the very
intense H$\alpha$\,
line was veiled. We favor values closer to the upper limit ($\> 2.7$)
in Table\,\ref{tab2}. The ratios of lines comparable in intensity
are in better agreement from different observations.
In the case of absolute fluxes, we used the highest values
obtained in the best  (but not perfect) conditions.
The same is true of  FWHM and  radial velocity shifts. We only present
the values from the higher resolution spectra. Generally the
resolution is not so relevant  in the case of line center
measurements. But we are
doubtful about instrumental line profiles and prefer  not to rely on them.
It should be noted, however, that in all the low resolution spectra
the blueshifts of the
lines are higher.

\begin{table*}
\caption{Measurements  of spectral lines}
\begin{tabular}{lccccc}
        \hline
        \noalign{\smallskip}
Emission & Equivalent Width &  Flux$^{\mathrm{a}}$($\times10^{-14}$)
& Relative &
FWHM$^{\mathrm{b}}$ & Line Center$^{\mathrm{c}}$ \\
   Line    & Range   (\AA)  &
({\rm{erg cm$^{\rm -2}$ s$^{\rm -1}$ }})    & Flux Range & \AA\, &
Shift (\AA)\\
   \noalign{\smallskip}
   \hline
   \noalign{\smallskip}
\ion{H$\epsilon$}    & -3--5 & 0.13 & 0.11--0.14 & -- & -- \\
\ion{H$\delta$}    & -6--12 & 0.3 & 0.23--0.34 & --  & --  \\
\ion{H$\gamma$}    & -16--18 &  0.67  & 0.41--0.57 & -- & -- \\
\ion{H$\beta$}     & -50--80 & 1.19  & 1.0 & 2.0 & -2.83  \\
\ion{H$\alpha$}    & -400--600 & 3.01 & 2.46--2.9 & --  & --  \\
\ion{He}{ii}~{$\lambda$\,4686\AA}& -30--44 & 1.09  & 0.7--0.92 & 1.84
& -2.68 \\
\ion{[O}{iii]} {$\lambda$\,5007\AA} & -2--3 & 0.037 & 0.03-0.035  & -- &--  \\
\ion{He}{ii}~{$\lambda$\,5411\AA} & -5--7  &  0.064  & 0.05--0.065 &
-- & --  \\
\ion{He}{i}~~{$\lambda$\,5876\AA} & -- & --  & $ <0.01$  & -- & --  \\
\ion{[N}{ii]}~{$\lambda$\,6584\AA} & --  & --  &$< 0.005$ & -- & --   \\
\ion{[S}{ii]}~~{$\lambda$\,6725\AA}  & -- & --  &$< 0.005$ & -- & --  \\
         \noalign{\smallskip}
        \hline
      \end{tabular}

     \label{slines}
\begin{list}{}{}
\item[$^{\mathrm{a}}$] flux measurements are based upon
the highest values obtained from SPM 2000 observations with a wide slit.
\item[$^{\mathrm{b}}$] as measured on the higher resolution INT spectrum.
\item[$^{\mathrm{c}}$] the shifts as measured on the  INT spectrum correspond to
a heliocentric velocity of -190$\pm$2\,km/sec.\\ the shifts measured on
lower resolution SPM spectra are systematically higher: of order
of -260\,km/sec
\end{list}
\label{tab2}
     \end{table*}

     \begin{figure*} 
    \resizebox{\hsize}{!}{\includegraphics[angle=-90]{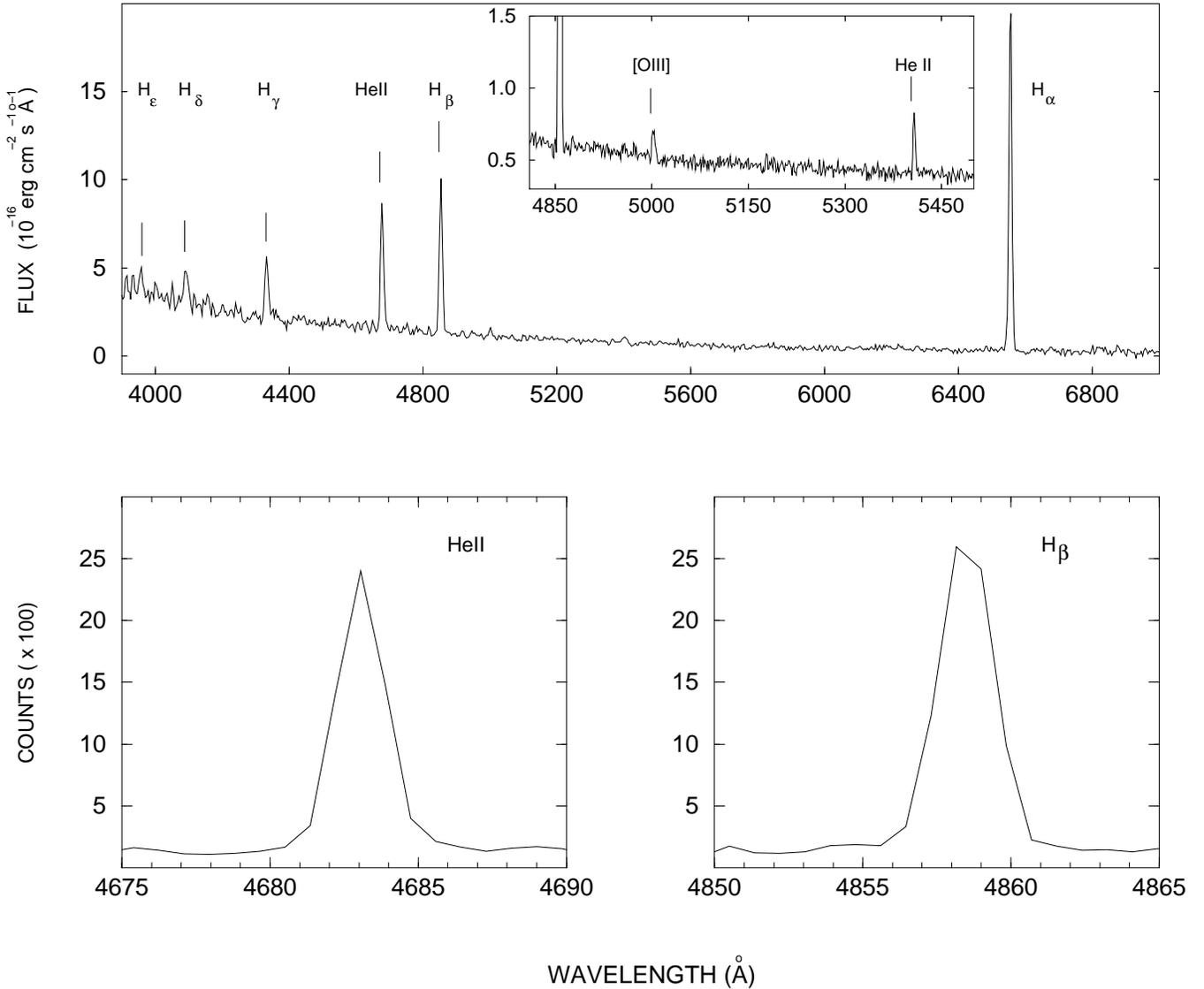}}
\hfill
\caption{The spectrum of  SBS\,1150+599A. In the upper panel  the single
spectrum obtained with a wider slit is presented. In the inset of the upper
panel the portion of the summed and averaged spectra is shown to outline the
presence of faint lines. In the lower panels the profiles of
\ion{He}{ii}~{$\lambda$\,4686\AA} and
H$\beta$ from the higher resolution spectrum are presented.}
\label{fig1}%
      \end{figure*}
%


\section {Nature of the object}

The optical spectra of \sbs\,  that we have at our
disposal
do not permit us to classify it unambiguously into any known  type of
object. Below we will discuss the most plausible options and will seek to
identify the nature of the object and model it.

\subsection {A Cataclysmic Variable ?}

Although \sbs\, was originally classified as a CV by the authors of the
SBS,
a careful examination of the spectra rejects such an interpretation completely.
CVs are close binary systems where stellar components are white an red dwarfs,
and the radiation is dominated by the outcome of accretion processes,
usually  trough disc accretion. CV systems are 
observed mostly in our vicinity (100--500 pc). The emission lines 
originate in accretion discs, they are broad due to the dispersion of 
velocities in the disc, and show radial velocity variations on short terms,
because of orbital motion in the binary (Warner \cite{war95}).   
Meanwhile, the emission lines of \sbs\, are narrow ($\leq2$\,\AA) and 
they lack any structure or radial
velocity changes on a nightly basis. Most importantly the lines show
persistent blueshift of  about 3\,\AA\,  at all epochs. A CV
with emission lines shifted to that extend should be a high inclination
system, thus in a time span of several hours (5--6 hours are half of the
longest period of CVs)  should demonstrate significant radial
velocity variations.

As mentioned above, in CVs one should expect much wider line profiles, due to
the presence of an accretion disc or accretion flow. Accreting matter is the
major source of light in CVs and even in magnetic CVs, where no disc is
formed, emission lines should contain much broader components,
variable with orbital phase (Schwope \cite{schwope}).
 
Finally, CVs with strong \ion{He}{ii} emission as a rule contain also
significant
neutral helium and show strong emission of
\ion{He}{i}\,$\lambda$\,4471, 5876\,\AA,
though they are not known to have any measurable [\ion{O}{iii}]
$\lambda5007$\,\AA\,
emission.

\subsection {A Symbiotic Star ?}

The spectrum of \sbs\, looks more like that of a  symbiotic star (SS) than
of a CV. SSs are binary systems  that contain a giant  secondary component
(at least for 80\% of them, the rest containing a Mira variable)
instead of a late type dwarf as in CVs. They
can be discovered at much larger distances, as far as in the
halo or in the Magellanic Clouds. Quite a few SSs are known to be halo objects.
They are
also known to contain a variety of nebular lines ([\ion{O}{iii}],
[\ion{Ne}{iii}],
   [\ion{Ne}{v}] and [\ion{Fe}{vii}]) not seen in
CVs.

For years, symbiotic stars were a collection of strange objects.
   However starting from 80's, two
catalogs of SSs and suspected objects were published (Allen \cite{alle84};
 Kenyon  \cite{ken86}) consolidating  a group of characteristics necessary for
   an object to be
classified as a symbiotic star. More recently in a new catalog of symbiotic
stars (Be{\l}czi\'nski et al. \cite{bel}) the classification criteria of symbiotics were refined,
based on a compromise between a large variety of features  displayed by more
than 180 objects classified as SSs.

The three basic characteristics that an object should fulfill in order to be
classified as a SS are:
\begin{itemize}
\item the presence of absorption lines of a late type giant;
\item the presence of emission lines of \ion{H}{i} and \ion{He}{i}
in combination either with lines from ions with ionization
   potential $\geq$\,35\,eV
   like [\ion{O}{iii}], or with a A--F type continuum with additional
   shell absorptions
   in hydrogen and helium lines;
\item the presence of the $\lambda$\,6825\,\AA\, emission feature, even if no
features of the cool star are observed.
\end{itemize}

\sbs\, apart from having high ionization emission lines does not satisfy
any other criteria established by  Be{\l}czi\'nski et al. (\cite{bel}).
Neither our spectra, nor available
IR surveys provide any evidence for the presence of late giant in the immediate
neighborhood of the object.

Another argument against a SS classification is that similarly to
CVs, SSs exhibit
complicated line profiles and strong variability. A broad component, in
addition to the narrow component, should be observed at least in
forbidden lines. Otherwise, if higher ionization lines are narrow and
featureless, one should observe central reversals or other structures
in the Balmer lines 
(see for example the atlas of high resolution line
profiles of SSs of Ivison et\,al. (\cite{ivi}).
Yet, the  lines of  \sbs\, do not show any deviation from
a single Gaussian profile or extended wings at the base in any of
the observed epochs.

Taking into account these arguments, we find it very unlikely that the
object in question is a symbiotic. However we cannot exclude this possibility
completely.
The uncertainty with the possible radial velocity variability  of the
lines leaves the question open until new extensive data can be collected.

\subsection {A Local Group Galaxy ?}

The object being blue shifted, if it were a galaxy it would
necessarily belong to the Local Group, implying that its distance
would be $<1.2$\,Mpc (van den Bergh \cite{bergh99}). From the observed H$\beta$
flux, we deduce that the total H$\beta$ luminosity would be
$<2\times10^3$\,\lsun.
This is much less than the typical H$\beta$ luminosities of \ion{H}{ii}
galaxies, which range between $10^{38}$ -- $10^{41}$ erg s$^{-1}$} 
(Salzer el al. \cite{sal89}). During the phase where O stars are
present, an ionizing star cluster with a total mass of $10^5$\,\msun\,
(typically
the mass of a globular cluster, Aguilar et al. \cite{agi88})
   produces about  $10^{51}$ -- $10^{52}$
ionizing photons per second, according to stellar population
synthesis models (Leitherer \& Heckman \cite{lehe95}). This implies that, if
such a cluster were powering our object, the surrounding gas would
absorb only about 1/1000 of the available ionizing photons. There is
no sign of presence of large quantities of dust in this object (from
the observed continuum  and from the H$\alpha$/H$\beta$ ratio) that would
absorb the radiation, so most of the cluster's ionizing radiation
would leak out. Qualitatively, this would explain why the observed
\ion{He}{ii}$\lambda$\,4686\,\AA/H$\beta$ ratio is so high
(in \ion{H}{ii} galaxies
it is at most equal to a
few percent, see e.g. Izotov \& Thuan \cite{izo98}) and why no line of
\ion{He}{i} is seen. However, the presence of
an \ion{He}{ii} emission line in an \ion{H}{ii} galaxy is generally
associated with a
broad \ion{He}{ii} feature produced by Wolf-Rayet stars (Izotov \& Thuan \cite{izo98}).
One may expect to see a narrow \ion{He}{ii} emission in a galaxy if it is due
to the ionization of gas by post-AGB stars (cf. Binette et\,al. \cite{bin94}).
But in that case, the expected equivalent width of the emission lines
would be much smaller than observed in our object and the color of the
stellar population would be red, not blue. In our spectra, we see no
signature of the presence of old stellar populations like the CN
feature ($\lambda$\,4150-4214\,\AA), or G band ($\lambda$\,4284-4318\,\AA),
or \ion{Mg}{i}
($\lambda$\,5156-5196\,\AA) that are found even in \ion{H}{ii} galaxies
(Raimann et al. \cite{rai00}). One should note however, that, if this
object were an \ion{H}{ii} galaxy, its metallicity would be extremely low in
order to produce such a small [\ion{O}{iii}]/H$\beta$ ratio, and this could
explain why the stellar metallic features would not be seen.

\begin{figure} 
    \resizebox{\hsize}{!}{\includegraphics{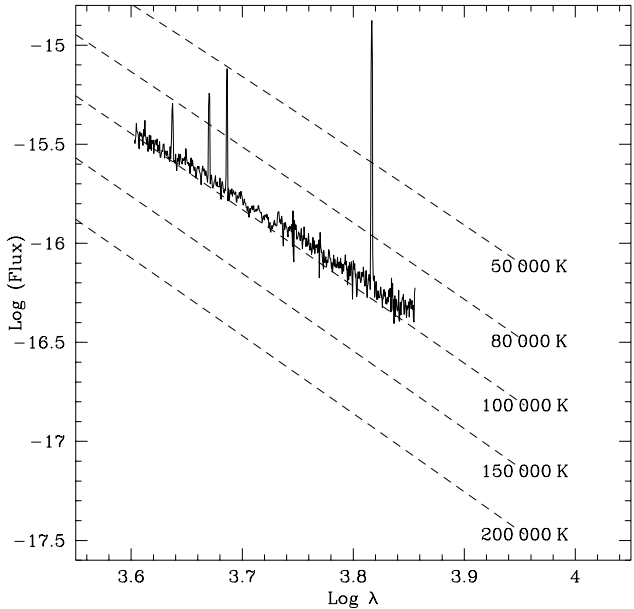}}
\hfill
\caption{The low resolution spectrum of  SBS\,1150+599A obtained at 
Cananea observatory is presented 
in order to show the steepness of the energy distribution in the continuum.
Blackbody curves of various temperatures are overplotted
for comparison.}
\label{fig2}%
      \end{figure}

Perhaps the most compelling argument against \sbs\ being an
 \ion{H}{ii} galaxy is the extremely blue nature of the continuum. It is
 much bluer than the continuum observed in 
blue compact galaxies (see e.g. Izotov \& Thuan (\cite{izo98}) and references 
therein). Fig. \ref{fig2} shows the 
observed continuum of \sbs\ in the log Flux -- log wavelength  plane, with blackbody
 functions of different temperatures superimposed on it. It clearly indicates 
that the continuum has a color temperature above 50\,000\,K. 
 Even active galactic nuclei do not show such a continuum in the 
 the $\lambda$\,4000--7000\,\AA\, wavelength range 
 (Morris \& Ward \cite{mm88}).

%

\subsection {A Halo Planetary Nebula ?}

Planetary nebulae are the product of the evolution of post-AGB
stars which, after having expelled their outer layers in stellar
winds, become sufficiently hot to ionize their gaseous envelope.
While the spectrum of our object is by no
means similar to that of a typical planetary nebula, since it shows
only one forbidden line between 4000 and 7000\,\AA, we may note that there
are examples of planetary nebulae in which the lines that are lacking
in the spectrum of \sbs\,  are very weak (Aller \& Czyzak \cite{alkr83}; 
Kingsburgh \& Barlow \cite{kiba94}). At first sight, our spectra can be
interpreted as pertaining to a planetary nebula with a hot exciting star
(to produce significant \ion{He}{ii} emission), that is density bounded (so
that no \ion{He}{i}, [\ion{N}{ii}] and [\ion{S}{ii}] lines are seen), and of
very low (or
perhaps extremely high) metallicity, so that the [\ion{O}{iii}]/H$\beta$ ratio
is very small, or in very high ionization state. 
For a PN  located in the Galactic
halo, one can reasonably eliminate the high metallicity option.
Note that observed heliocentric velocity of 190--260 km/sec  
(see Table 2) is indeed compatible with \sbs\ belonging to the 
Galactic halo population (Beers \& Sommer-Larsen \cite{bee95};  
Beers et al. \cite{bee00}).

The observed expansion velocity determined from the
HWHM is 40\,km~sec$^{-1}$ (taking a FWHM of 2\,\AA\, and
an instrumental width of 1.5\,\AA). This is rather high for a PN
   if one compares with the values listed in the catalogue of expansion
   velocities of
Galactic planetary nebulae by Weinberger (\cite{wei89}) and in the paper
by Dopita et\,al. (\cite{dop85}) on PNe in the Magellanic Clouds, but
there are well known PNe which do show similar
expansion velocities, for example PNe with [WR]
type nuclei (Pe\~na et al. \cite{pe01}).

In the following section, we further elaborate on the PN hypothesis,
   with the tools developed for the
analysis of this type of objects.

\section {A model for SBS\,1150+599A as a planetary nebula.}

\subsection {Photoionization computations}

Given the small number of observational constraints, we adopt a very
simple model of a planetary nebula which consists of a homogeneous
gaseous sphere surrounding a hot star radiating as a blackbody. In
order to put some limits on the stellar temperature and oxygen
abundances, we construct series of photoionization models in which we
specify the stellar temperature, T$_*$, the number of photons $Q$
emitted per second by the star at energies above 13.6\,eV, the gas
density $n$, and the filling
factor $\epsilon$ (both assumed constant within the nebula).
Each
series consists of 12 models with metallicities ranging from 
$10^{-4}$\,solar to
30\,solar (the high abundance end is shown only for didactic purposes
since, as mentioned before, we do not expect a high metallicity
object in the halo). Oxygen
is the reference element (with the solar abundance in units of
$\log \ion{O}/\ion{H}+12$ being taken equal to 8.93 from Anders \& Grevesse
(\cite{angr89})).

The abundances of the other elements relative to oxygen follow the prescription of
Mc\,Gaugh (\cite{mc91}). Each model is constructed with the
photoionization code PHOTO as described in Stasi\'nska \& Leitherer (\cite{stle96}).
The computations are performed starting from the center and are
stopped when the equivalent width of the H$\beta$ line, EW(H$\beta$),
becomes equal to 130\,\AA. This value takes into account the fact
that the spectra do not cover the object completely (we have checked
that changing
EW(H$\beta$) by 50\% will not change fundamentally our results).
We do not have to consider all possible combinations of
$Q$, $n$ and $\epsilon$ since for a fixed effective temperature and
a given condition imposed on EW(H$\beta$),
the nebular ionization structure depends only on $Q\,n\,\epsilon^2$.

     \begin{figure*}
    \resizebox{17cm}{!}{\includegraphics{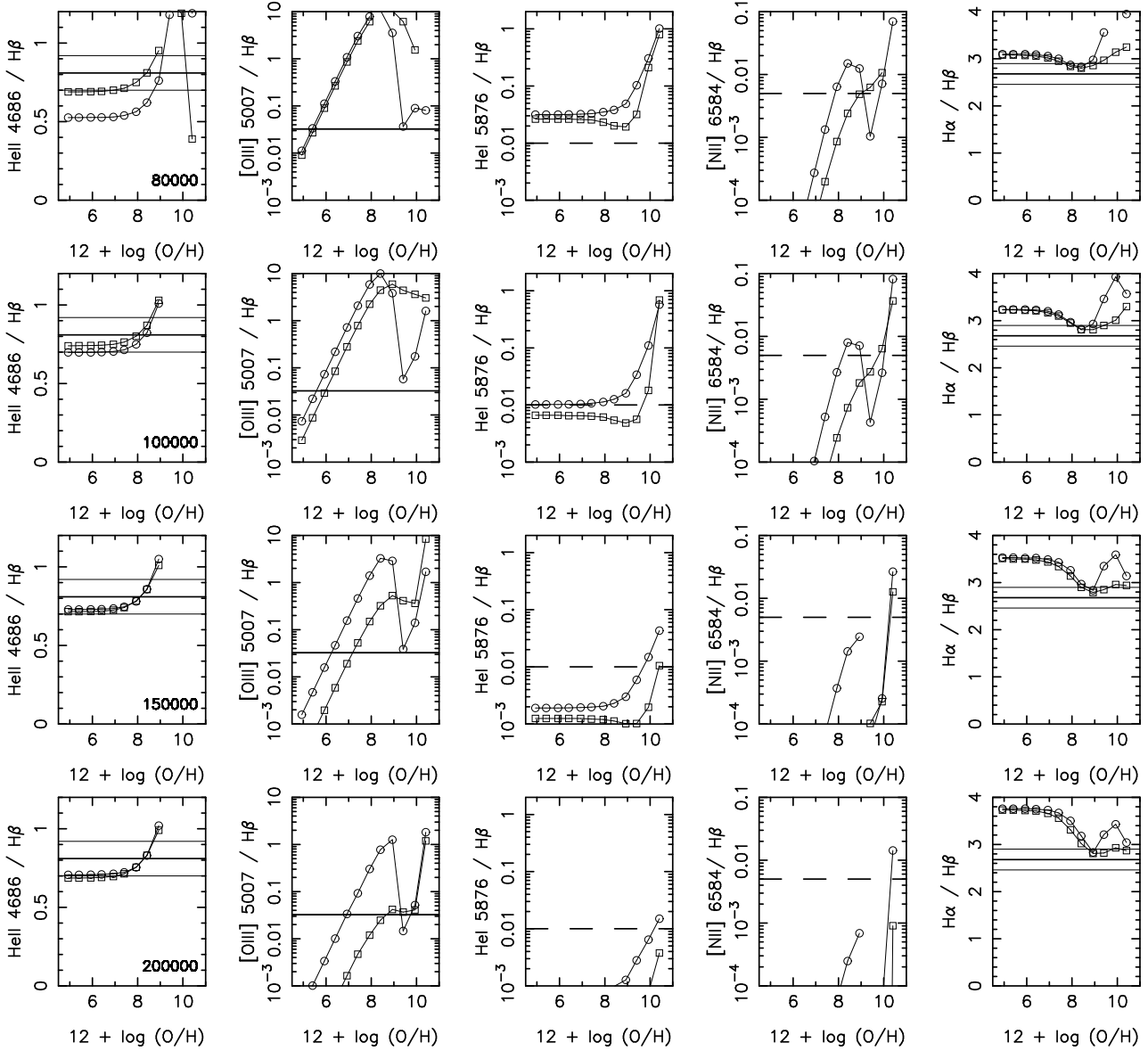}}
\caption[]{Sequences of photoionization models with varying O/H. Each
row of panels corresponds to a different value of T$_*$
as indicated on the leftmost panel. Circles: models with $n$=
100\,cm$^{-3}$;
squares: models with $n$=
$10^4$\,cm$^{-3}$. All of the models are constructed using
$Q=3\times10^{47}$\,ph\,s$^{-1}$, and
$\epsilon =0.2$, and are density bounded in order to give
   EW(H$\beta$) = 130\,\AA\ (see text).
   The observed line ratios are indicated by horizontal lines (dashed
   lines in the case of upper limits, thick continuous lines for measured
   values, thin lines for the error bars.}

\label{fig3}
      \end{figure*}

Figure \ref{fig3} shows the results of our models, computed with
$Q=3\times10^{47}$\,ph\,s$^{-1}$, a nebular filling factor
$\epsilon =0.2$ (values that are typical
for planetary nebulae), and two values of $n$:  100\,cm$^{-3}$ (circles) and
$10^4$\,cm$^{-3}$ (squares), for different values of T$_*$. The first row of
panels corresponds to T$_*=80\,000$\,K, the second to 100\,000\,K,
the third to
150\,000\,K, and the fourth to 200\,000\,K. From left to right, the panels
show the behavior of \ion{He}{ii}\,$\lambda$\,4686\,\AA/H$\beta$,
[\ion{O}{iii}]\,$\lambda$\,5007\,\AA/H$\beta$,
\ion{He}{i}\,$\lambda$\,5876\,\AA/H$\beta$,
[\ion{N}{ii}]\,$\lambda$\,6584\,\AA/H$\beta$, and H$\alpha$/H$\beta$
as a function of $\log(\ion{O}/\ion{H})+12$. The
observational constraints are shown by a thick horizontal line if the
relevant line is measured (the thin horizontal lines show the error
bars) or a dashed line if only an upper limit is available.

All the models show qualitatively a similar behavior. For a given
value of T$_*$, the effect of increasing the density at a given
\ion{O}/\ion{H}\,
is to increase
the ionization parameter, 
so that the ratios \ion{He}{i}\,$\lambda$\,5876\,\AA,
[\ion{N}{ii}] and
[\ion{O}{iii}] decrease.  As \ion{O}/\ion{H}\, increases, cooling becomes
more important and the electron temperature drops.
This produces an increase in the
\ion{He}{ii}\,$\lambda$\,4686\,\AA/H$\beta$ and
\ion{He}{i}\,$\lambda$\,5876\,\AA/H$\beta$ ratios,
due to the temperature
dependence of the emissivities of these lines. On the other hand,
[\ion{O}{iii}]/H$\beta$ and
[\ion{N}{ii}]/H$\beta$ first increase (abundance effect) then decrease
(cooling is being gradually shifted from optical to
infrared lines). At very high
metallicities, these ratios rise again, because the [\ion{O}{iii}]
and [\ion{N}{ii}]
lines are now mainly produced by recombinations, due to the very low
electron temperature. At metallicies less than solar, the values of
H$\alpha$/H$\beta$ in our models decrease with increasing \ion{O}/\ion{H},
while the
recombination value (Brocklehurst \cite{bro71}) would increase. This is due to
the fact that, at high electron temperature, collisional excitation 
contributes to the H$\alpha$ emission.
Clearly, our models show that there is quite a range of input
parameters that are compatible with our spectroscopic observational constraints
(if one dismisses the H$\alpha$/H$\beta$ ratio which poses a problem to
which we will come back below). While models with T$_*$ $=$ 80\,000\,K are not compatible
with the lack of detection of \ion{He}{i}\,$\lambda$\,5876\,\AA, models with
T$_*$ between
100\,000\,K and 200\,000\,K are consistent with the observed constraints
in a range of metallicities that depend essentially on the value of the
ionization parameter (which, in our parametrisation of the problem,
is only dependent upon the density). For each value of $n$, the value of
O/H which matches the observed [\ion{O}{iii}]/H$\beta$ ratio increases
with T$_*$, since more and more oxygen ions are found in ionization
stages higher than O$^{++}$. At a given T$_*$, the solution for O/H
increases with $n$ for the same reason. Clearly, one needs more
observational constraints to determine O/H and  T$_*$.

Note that, as seen in Fig. \ref{fig2} ,  
the observed continuum is compatible with any value of 
T$_*$ above 70\,000\,K and does not provide additional constraint.

\subsection {The problem of H$\alpha$/H$\beta$}

It is not satisfactory that the ratio  produced by the
models is much higher than observed. Our computations assume case B.
Case A would better apply in the case of our object which is extremely
density bounded, but the effect on  H$\alpha$/H$\beta$ would remain
  virtually unchanged.
In our models, collisional excitation of H$\alpha$ strongly
increases H$\alpha$/H$\beta$  with respect to the recombination value.
In order to avoid collisional excitation, one needs to lower the 
electron temperature. This can be achieved by changing the geometry so 
as to increase the fraction of trace neutral hydrogen 
and, thus, enhance H Ly$_\alpha$ cooling. 
We have tried to obtain such models, but they required
very unrealistic conditions.

Since we are not absolutely certain about the calibration of the
H$\alpha$/H$\beta$ ratio in our spectra, we intend to obtain additional
spectroscopy of \sbs\, in order to check  the true value of
this ratio. We note, however, that in the lists of PNe published by
Aller \& Keyes (\cite{all87})  or Kingsburgh \& Barlow (\cite{kiba94}), there are other
planetary nebulae with H$\alpha$/H$\beta$ ratios similar to the one we
find here for \sbs. The authors of those papers do not comment on
their finding, except to say that in those cases the reddening to the
object was considered zero.

Interestingly, in some respects, our object is similar to the one found by
Skillman et al. (\cite{ski89}) in the dwarf irregular galaxy Leo A, which shows relatively strong
\ion{He}{ii}\,$\lambda$\,4686\,\AA, no detectable
[\ion{O}{ii}]\,$\lambda$\,3727\,\AA and no detectable continuum emission
and which they interpret as being a PN. Their object shows an
H$\alpha$/H$\beta$ of 2.3, which they tentatively attribute to scattering
by dust near the PN.

There are astrophysical objects (old novae, cataclysmic variables, Be stars)
where the H$\alpha$/H$\beta$ is
notoriously far below the recombination value, reaching values of
the order of one.  However,
the conditions invoked to explain the flat observed Balmer decrements,
local thermal equilibrium (Williams \cite{will80}) or stimulated emission by a
strong radiation field (Elitzur et al. \cite{eli83}; Williams \& Shipman \cite{will88})
require a high density, which is excluded in the
 case of our object, since it shows the forbidden line of [\ion{O}{iii}].

\subsection {Distance and structural properties of SBS\,1150+599A}

We now make use of other properties of \sbs, to further
constrain the models. From the observed angular radius and H$\beta$
   flux
(assuming that interstellar extinction is negligible, which is likely
from the observed value of H$\alpha$/H$\beta$ and from the high
Galactic latitude of the object)  we can compute the distance of
the object assuming a certain mass and filling factor of the gas
emitting in H$\beta$. Since our object is density bounded, such a
procedure (the Shklovsky method) provides a reasonable estimate of
the true distance to the object. We have
$$d=0.14\,F_{\mathrm {H\beta}}\,^{-0.2}\,\theta^{-0.6}\, M\mathrm{_{neb}}\,
^{0.4}\,\epsilon^{-0.2}$$
where $d$ is the distance in kpc, $F_\mathrm {H\beta}$ is the H$\beta$
flux in  erg\,cm$^{-2}$\,s$^{-1}$, $\theta$\,  is the angular radius in arcsec
and $M_\mathrm {neb}$ is the nebular mass in solar masses. We adopt an
angular radius of 5$\arcsec$, corresponding to the spatial extension 
of the emission 
lines in our spectra and a total H$\beta$ flux  $F_\mathrm
{H\beta}$= 2.4$\times10^{-14}$ erg\,cm$^{-2}$\,s$^{-1}$, assuming that 
the nebula is round.

Taking $M_\mathrm {neb}=0.2$\msun\, and $\epsilon=0.2$  gives $d=20.4$\,kpc
which is
   within acceptable distances for a halo object (Zaritsky 1999).
Taking $M_\mathrm {neb}=0.1$\msun\,  rather than $0.2$\msun\,  gives
$d=15.5$\,kpc, taking
$\epsilon=1$ rather than 0.2 gives $d=14.8$\,kpc.

Adopting as a working hypothesis the values $M_\mathrm {neb}=0.2$\,\msun\,
and  ~$\epsilon=0.2$,
we obtain a nebular radius R$_\mathrm{neb}$ of 0.5\,pc and
a gas density of
$60\,\mathrm{cm}^{-3}$. The values of R$_\mathrm{neb}$  and $n$
obtained using various combinations of $M_\mathrm {neb}$ and $\epsilon$,
are reported in Table \ref{modtab}.

\begin{table}[th]
\caption{Derivation of the structural parameters of \sbs }
\begin{tabular}{lccc}
        \noalign{\smallskip}
        \hline
        \noalign{\smallskip}
   $M_\mathrm {neb}$ [\msun]  & 0.2 & 0.2 & 0.1  \\
    $\epsilon$ & 1 & 0.2 & 0.2 \\
    \noalign{\smallskip}
     \hline
     \noalign{\smallskip}
    $d~[kpc]$ & 14.8 & 20.4 & 15.5 \\
   $R_\mathrm {neb}$ [pc] & .36 & .50 & .40 \\
   $n$ [cm$^{-3}]$ & 30 & 59 & 57 \\
   $t_\mathrm {exp}$ [yr] & 9100 & 12500 & 10000 \\
      \noalign{\smallskip}
   \hline
        \noalign{\smallskip}
   $T_{*}$ [K] & $Q^a$  & $Q$  &$Q$  \\
   log O/H + 12  &  $M_{*} (Q)^{b}$  & $M_{*} (Q)$  &
   $M_{*} (Q)$ ) \\
     &  $M_{*} (T_{*})^{b}$ & $M_{*} (T_{*})$ &
     $M_{*} (T_{*})$  \\
\hline
\noalign{\smallskip}
100\,000 & 625 & 1220 & 593 \\
5.8 & -- & --  & -- \\
    & 58.5 & 57.5 & .58 \\
    \noalign{\medskip}
     150\,000 & 1450 & 2850 & 1370 \\
6.2 & .59 & .57 & .59 \\
    & .70 & .58 & .59 \\
    \noalign{\medskip}
    200\,000 & 2530 & 4980 & 2400 \\
7. & .585 & -- & .58 \\
    & --  & --  & --  \\

         \noalign{\smallskip}
        \hline
\noalign{\smallskip}
$^a$  in $10^{44}$ ph $s^{-1}$\\
$^{b}$ in \msun
      \end{tabular}
     \label{modtab}
     \end{table}

It is easily shown that, for a star of given T$_*$, the observed EW(H$\beta$)
can be obtained by a whole family of models sharing the same  value of
   ($M_\mathrm {neb}$$n$/$Q$) (for a given electron temperature). From the
   models we have computed, and which return the value of  $M_\mathrm
   {neb}$ that corresponds to EW(H$\beta$)=130\,\AA, we can readily
infer the value of $Q$ that corresponds to a chosen
   $M_\mathrm {neb}$.
These values,  which produce acceptable models,
 are reported in Table \ref{modtab}, for different choices
of T$_*$. One sees that the values
of $Q$ are all within a factor 5 of the input value in our models
presented in Fig. \ref{fig3}. The densities reported in
Table \ref{modtab}\,  for \sbs\,  are of the order of 50\,cm$^{-3}$.
Therefore, of the series of models presented in Fig. \ref{fig3}, those
with $n$=10$^{4}$\,cm$^{-3}$ are not relevant for our object but those 
with $n$=100\,cm$^{-3}$ have an ionization parameter roughly 
compatible with the results from  Table \ref{modtab}. We can then infer an
approximate oxygen abundance for \sbs\, by finding where the
sequences represented by circles meet the observed value of
[\ion{O}{iii}]\,$\lambda$\,5007\,\AA/H$\beta$ in Fig.
\ref{fig3}. These abundances are reported in Table \ref{modtab}
below the corresponding T$_*$. The accepted values still span a large
range, but oxygen abundances over one fiftieth of solar are clearly
eliminated. This makes of \sbs\,
by far the most oxygen--poor PN known.

     \begin{figure*}
    \resizebox{19cm}{!}{\includegraphics{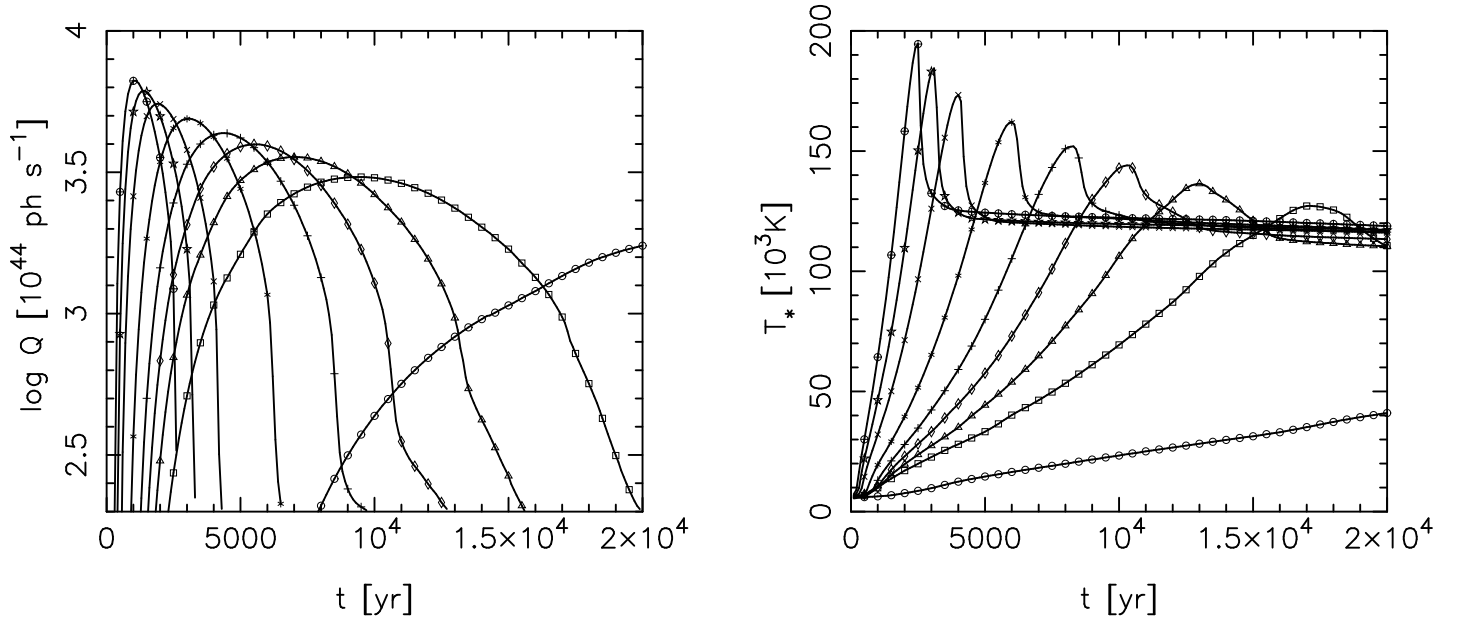}}
\caption[]{The values of $Q$ and T$_*$ as a function of time
interpolated from the grid of
post-AGB tracks by Bl\"ocker (\cite{blo95}) for central star masses from 0.56
to 0.64\msun by intervals of 0.01\msun. The evolution is the slowest
for the lowest stellar masses.}

\label{fig4}
      \end{figure*}

By using a grid of stellar evolution tracks for post-AGB stars (we use
the one computed by Bl\"ocker (\cite{blo95}) for hydrogen burning PN nuclei),
it is possible to go one step further, if one assumes
that the transition time
between the PN ejection and the beginning of the
evolution of the nucleus off the AGB  is zero. We first estimate
the expansion time, from the formula
   $t_\mathrm {exp}$=$R_\mathrm {neb}$/$v_\mathrm {exp}$. The derived number
   depends on the assumptions on  $M_\mathrm {neb}$ and
   $\epsilon$ (see Table  \ref{modtab}), but is always around 10\,000~yr.
   Then, from Fig.  \ref{fig4}, which has been obtained by
   interpolating the grid of Bl\"ocker (1995) on a finer grid in star
   masses, we derive $M_{*} (Q)$ and $M_{*} (T_{*})$, the values of the
   PN nuclei masses that correspond to $Q$ and $T_{*}$ respectively.
   These values are reported in Table  \ref{modtab}. If the central
   star evolved exactly as predicted by  Bl\"ocker (\cite{blo95}) and if \sbs\,
   were perfectly represented by our toy model, the true central star
   mass would be obtained for the value of  $T_{*}$  which makes $M_{*}
   (Q)$ and $M_{*} (T_{*})$  equal. From an inspection of Table
   \ref{modtab}, one finds for the central star of \sbs\,  a temperature
   of $\sim$ 150\,000~K
   and a mass of $\sim$ 0.58--0.59~M\sun.

   These numbers are not to be taken too literally, of course. Our model
   is extremely simplistic. Hydrodynamical computations (Bobrowsky \&
   Zipoy \cite{boz89}; Mellema \cite{mel94})
   have warned against a blind use of expansion times and we know that
   PNe are seldom spherical with a uniform density. Besides, it has been shown
   empirically by
studying samples of planetary nebulae in various galaxies, that the
central stars of planetary nebulae evolve differently according to
the galactic context (Stasi\'nska et al. \cite{sta98}; Dopita et al. \cite{dop97}).

Nevertheless, the fact that, with our simple approach, we are able to
give a reasonable picture of \sbs\, is extremely encouraging. As a
consequence, we favor for \sbs\, an oxygen abundance of log O/H+12 $\sim$
5.8--6.8.

\subsection {The helium abundance in \sbs}

The observed \ion{He}{ii}\,$\lambda$\,4686\,\AA/H$\beta$  
leads to a He$^{++}$/H ratio of 
$0.072\pm0.010$. The observed limit on \ion{He}{i}\,$\lambda$\,5876\,\AA\
 leads to He$^{+}$/H $< 0.009$. Unfortunately, due to the quality of our data, 
 these estimates 
are not sufficiently accurate to provide any useful indication of the 
primordial helium abundance. The only thing which can be said at this 
stage is that the helium abundance we are finding is consistent with 
our interpretation that \sbs\ is an extremely oxygen poor PN.

\subsection {Need for further observations}

     \begin{figure*}
    \resizebox{17cm}{!}{\includegraphics{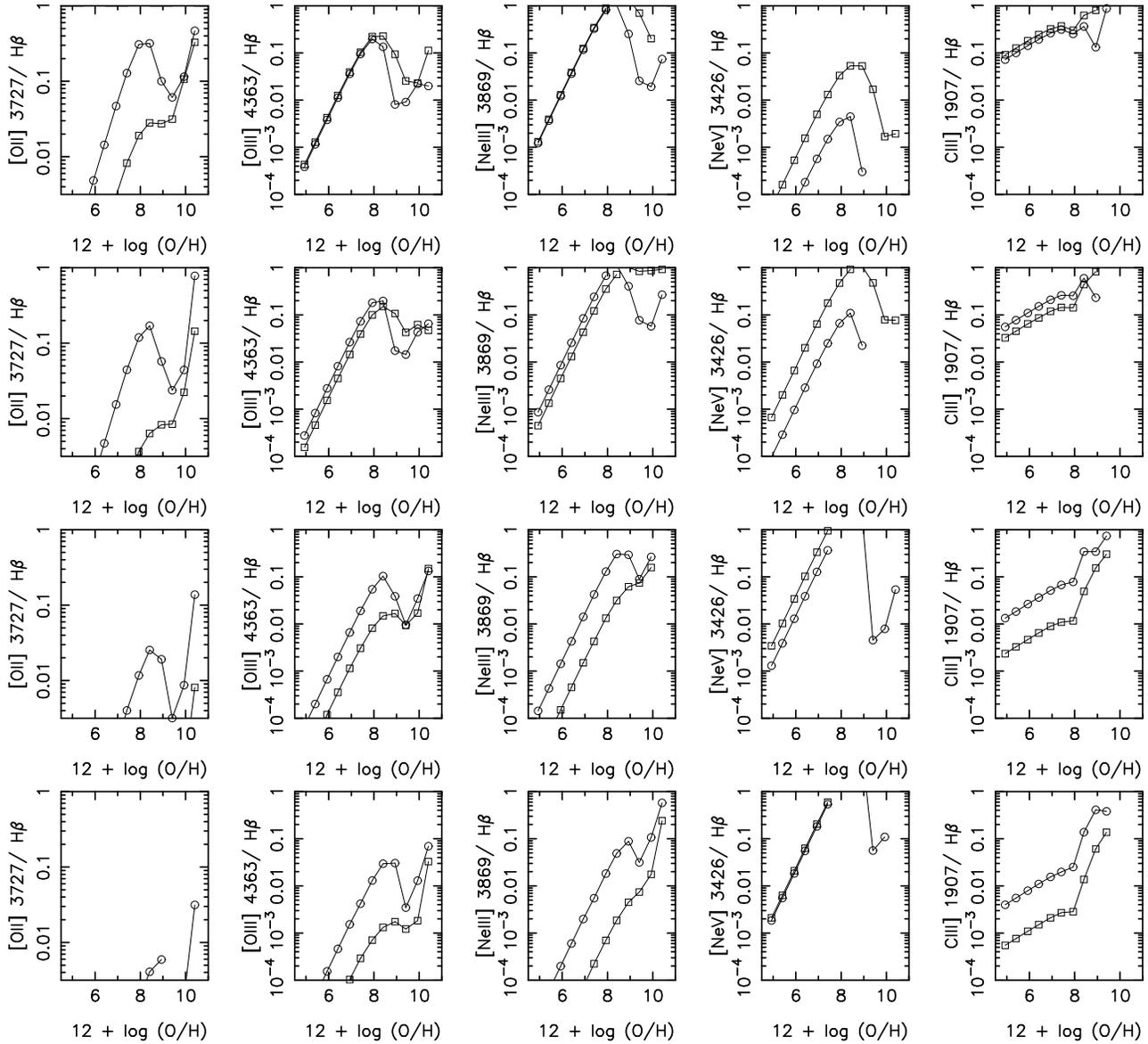}}
\caption[]{Predicted values of some line ratios for the same models as shown in Fig\ref{fig3}.}

\label{fig5}
      \end{figure*}

One way to better constrain the oxygen abundance (and to test the
hypothesis of mixing, see Sect. 6) in \sbs\,  would be through deeper
spectroscopy in the optical, which should reveal the weak [\ion{Ne}{iii}]
$\lambda$\,3869\,\AA, [\ion{Ne}{v}]$\lambda$\,3426\,\AA\, lines, 
and possibly also
the lines from the low
excitation e.g., [\ion{O}{ii}]$\lambda$\,3727\,\AA,
[\ion{N}{ii}]$\lambda$\,6584\,\AA.
Our models
show that even the [\ion{O}{iii}]$\lambda$\,4363\,\AA\,
line could be measurable. This would considerably improve the quality
of the oxygen
abundance determination. Ultraviolet spectra would bear important
information.  For example, they would place  limits on the
carbon abundance and on the ionization structure of the nebula
through the [\ion{O}{iv}]$\lambda$\,1394\,\AA\,
line. These lines are expected
to be measurable, despite the low metallicity of the object, due to the
high electron temperature which boosts the collisionally excited
ultraviolet lines with respect to the optical ones. A direct
observational constraint on the nebular density would be very
valuable too. The optical doublets which are used for that ([\ion{S}{ii}]
$\lambda$\,6716, 6731\,\AA\, and [\ion{A}{iv}]$\lambda$\,4711, 4740\,\AA)
would be probably
too weak to be
measured, but one may hope to measure the
[\ion{C}{iii}]$\lambda$\,1907, 1909\,\AA\,
doublet in the
ultraviolet.

We show in Fig.\ref{fig5} the predicted intensities 
relative to H$\beta$ of some important lines, for the same models as 
in Fig.\ref{fig3}.  These predictions are purely indicative and are 
given only to estimate the feasibility of future observations.  For 
example the intensity of [\ion{Ne}{v}]$\lambda$\,3426\,\AA\, strongly 
depends on the assumed spectral energy distribution (here a simple 
blackbody was used).  That of [\ion{C}{iii}]$\lambda$\,1907, 
1909\,\AA\ is probably larger than shown here since C/O ratios in PNe 
tend to be larger (Henry et al. \cite{hkb}) than given by the prescription 
of Mc\,Gaugh (\cite{mc91}) used in these exploratory models.


\section{Discussion}

Our main conclusion is that, most probably, \sbs\, is a halo
planetary nebula. With the nomenclature of the Strasbourg-ESO  
catalogue of planetary nebulae (Acker at el. \cite{ack92}) it becomes PN G135.9+55.9. 
Our analysis favors an oxygen abundance lower than 1/100 and probably around 1/500 
of solar. 
This would make it the most oxygen--poor PN known so
far. Among the dozen of known Galactic halo PNe
(Conlon et\,al. \cite{con93};
Napiwotzki et\,al. \cite{nap94}; Howard et\,al. \cite{hwd97} and references therein; 
Jacoby et al. \cite{jac97}),
  the most oxygen--poor is K\,648
with a $\log\ion{O}/\ion{H} + 12$ estimated at 7.61.

How does this fit with our knowledge of the formation and evolution
of planetary nebulae and on the evolution of the Galactic halo?

It is obviously not surprising to find a low metallicity object in
the halo, hundreds of stars have been discovered with metallicities
(as referred to the iron abundance) [\ion{Fe}/\ion{H}]$ <-2$\,dex,
the lowest ones
reaching $-4$\,dex (see e.g. references in Beers \cite{bee99}; Norris \cite{nor99}).
Oxygen abundances, however, have been determined 
in only a fraction of these metal-poor stars (Laird \cite{lia99}). The lowest 
oxygen abundances measured in stars are around 1/200 solar
(Boesgaard et\,al. \cite{boe99}; Israelian et\,al. \cite{isr01}). It should be remembered,
though, that oxygen abundance measurements in halo stars are
difficult and uncertain (see the discussion by McWilliam \cite{mcw97}; Hill 
\cite{h01} and the papers edited by Barbuy \cite{b01}). Our
determination of the oxygen abundance in \sbs\,  is not very
accurate either, and it would be important to constrain it
better with  appropriate observations. Indeed, if the oxygen abundance
in \sbs\,  reflects the oxygen content of the material out of
which the progenitor star was formed, as is expected from theoretical
models of AGB stars (Forestini \& Charbonnel \cite{foch97}), a proper
determination of this abundance would give interesting clues to
the understanding of the formation of the Galactic halo and of the
first stars.

There is, however, the possibility that the oxygen abundance in \sbs\,
has been modified from the pristine oxygen abundance by
mixing processes in the progenitor star. Indeed, differences in
abundance ratios from star to star in globular clusters giants seem
to indicate that in those stars deep mixing has brought to the
surface the products of proton capture in the hydrogen burning shell.
(Ivans et\,al. \cite{iva99}; Weiss et al. \cite{wei00} and references therein)
However, this is not seen in field stars (Gratton et\,al. \cite{gr00};
 Charbonnel \& Palacios \cite{cp01}).
Usually, abundance ratios observed in PNe do not indicate that oxygen 
has been depleted in the progenitor by mixing (Richer et\,al. \cite{ric98}). 
There is however the case of the halo planetary nebula BB-1 which shows a 
\ion{Ne}/\ion{O}\, ratio
  much higher than the remaining
planetary nebulae (Howard et\,al. \cite{hwd97} and references therein),
and for which destruction
of \element[][16]{O} in
the \ion{ON}\, cycle has been suggested.

If the progenitor of \sbs\,  was indeed one of the most metal
poor objects known in the galaxy, could it have produced a planetary
nebula that we see now? With such a low metallicity, the star must be
older than the most metal-poor globular clusters, i.e. about 14\,Myr
(VandenBerg \cite{vdb99}) and from stellar evolution models (Charbonnel et\,al.
\cite{cha99}) we derive that the initial stellar mass would be $0.8 - 1.0$~\msun.
If we adopt the initial-final mass relation of Weideman (\cite{wei87}) or
even that of Richer et\,al. (\cite{ric97}), the mass of the planetary nebula
nucleus would be around $0.55 - 0.555$~\msun. According to the post-AGB
stellar evolution models of Bl\"ocker (\cite{blo95} and references therein), the
evolution of the star off the AGB would be so slow that the ejected
material would have been totally dispersed before being ionized.
There are however several possibilities to reconcile the low
metallicity of the object and the fact that it has produced a
planetary nebula visible now.
First, the star could have been formed from infalling material,
according to a scenario discussed by Carney (\cite{car99}). Then its age
would be smaller than that derived for the bulk of metal poor
globular clusters, and its mass larger than 1~\msun. Second, there
are many uncertainties in the derivation of the initial-final mass
relation (Weideman \cite{wei87}; Bragaglia et\,al. \cite{bra95}; Richer et\,al. \cite{ric97})
and it is possible that at low metallicities the final mass would be
larger, for example due to decreased mass loss. One cannot exclude
some intrinsic scatter in the initial-final mass relation, for
example due to close binary stellar evolution. Third, the stellar
post-AGB evolution itself can be questioned. For example, Kato \&
Hachisu (\cite{kaha93}) argue that computations using OPAL
opacities drastically shorten the evolutionary time scale.

For all of these reasons, then, our interpretation that \sbs\
is an extremely oxygen-poor halo PN does make sense.
Conversely, a more detailed study of this object would provide
valuable constraints on the evolution intermediate mass stars of low
metallicity and on the chemical evolution of the galaxy.

Is \sbs\,  unique?
If we take a total enclosed mass for our Galaxy (within 200 kpc) of
1.4$\times10^{12}$\msun, a mass to light ratio of 100 in solar units
(Zaritsky \cite{zar99}), and if we adopt a PN production rate of $50\times10^{-8}$
PN/\lsun  (Ciardullo \cite{cia95}), we arrive to a total of 7000 PNe in the Galaxy.
 From the observed metallicity distribution of
metal poor stars (Norris \cite{nor99}), one does not expect to find a large number of
such objects in the Galactic halo. However, the criteria for
discovering PNe in objective prism or imaging surveys
are based on the strength of the [\ion{O}{iii}] line. \sbs\, was
probably first discarded as a planetary nebula
because of its very weak [\ion{O}{iii}] line and because of the absence of
other emission lines usually seen in planetary nebulae. In this
respect, it might prove rewarding to reexamine the spectra of the
objects declared as cataclysmic variables or emission line stars in
objective prism surveys.

\begin{acknowledgements}

GT acknowledges CONACyT grant 25454-E and DGAPA grant IN113599. GT thanks DAEC for
hospitality and financial support. VHC acknowledges CONACyT research
grant 28499-E. The authors are grateful to C. Boisson, M. Richer,
M. Spite, F. Spite, C. Charbonnel, M. Pe\~na, J. Miko{\l}ajewska
for discussions, and to the referee J.A. de Freitas Pacheco 
for his valuable comments.
\end{acknowledgements}

\end{document}